\documentclass[preprint,showpacs,preprintnumbers,amsmath,amssymb]{revtex4}

\usepackage{graphicx}% Include figure files
\usepackage{dcolumn}% Align table columns on decimal point
\usepackage{bm}% bold math

\begin{document}

\title{\bf Resolving signals in the LISA data}

\author{Andrzej Kr\'olak}
\email{krolak@jpl.nasa.gov}
\altaffiliation [Also at: ]{Institute of Mathematics,
Polish Academy of Sciences, Warsaw, Poland}
\affiliation{Jet Propulsion Laboratory, California Institute of
  Technology, Pasadena, CA 91109}

\author{Massimo Tinto}
\email{Massimo.Tinto@jpl.nasa.gov}
\altaffiliation [Also at: ]{Space Radiation Laboratory, California
  Institute of Technology, Pasadena, CA 91125}
\affiliation{Jet Propulsion Laboratory, California Institute of
  Technology, Pasadena, CA 91109}

\date{\today}

\begin{abstract}
  We estimate the upper frequency cutoff of the galactic white dwarf
  binaries gravitational wave background that will be observable by
  the LISA detector. This is done by including the modulation of the
  gravitational wave signal due the motion of the detector around the
  Sun. We find this frequency cutoff to be equal to $10^{-3.0}$Hz, a
  factor of 2 smaller than the values previously derived. This implies
  an increase in the number of resolvable signals in the
  LISA band by a factor of about 4.

  Our theoretical derivation is complemented by a numerical
  simulation, which shows that by using the maximum likelihood
  estimation technique it is possible to accurately estimate the
  parameters of the resolvable signals and then remove them from the
  LISA data.
\end{abstract}

\pacs{95.55.Ym, 04.80.Nn,95.75.Pq, 97.60.Gb}

\maketitle

The gravitational waves from short-period binary systems
containing white dwarfs and neutron stars are signals guaranteed
to be observed by the Laser Interferometer Space Antenna (LISA)
mission.  Recent surveys indicate that there exist about twenty of
such systems that are emitting gravitational radiation of
frequency falling into the LISA band.  Population studies have
also shown that the number of such sources will be so large to
produce a stochastic background that will lie significantly above
the LISA instrumental noise in the low-part of its frequency band.
It has been shown in the literature (see \cite{NYP00} for a recent
study and \cite{HBW90,EIS87} for earlier investigations) that
these sources will be dominated by detached white dwarf - white
dwarf (wd-wd) binaries, with $1.1 \ \times 10^8$ of such systems
in our Galaxy. The detached wd-wd binaries evolve by
gravitational-radiation reaction and the number of such sources
rapidly decreases with increasing orbital frequency. It is
therefore expected that above a certain limit frequency one will
be able to resolve individual signals.

In this communication we calculate this limit frequency by
including the effect of the signal modulation induced by the
motion of LISA around the Sun. Earlier derivations assumed that
one would be able to resolve only one signal per frequency bin.
However, the modulation of the signal depends on the position of
the source in the sky and therefore, by using matched-filtering
technique, it is possible to resolve signals with same frequency
but incoming from different directions.

The number distribution, $\frac{dN}{df}$, of the detached wd - wd
binaries has been estimated in \cite{NYP00} to be equal to

\begin{equation}\label{DF}
\frac{dN}{df} = 1 \mbox{yr}
\left(\frac{f}{f_o}\right)^{-\frac{11}{3}} \ ,
\end{equation}
where yr is expressed in seconds, and $f_o = 10^{-2.8} \ $Hz. Thus, for an
observation time of $n$ years, the number of sources in one frequency
bin,$N_s$, can be approximated by the following formula
\begin{equation} N_s =
\left(\frac{f}{f_o}\right)^{-\frac{11}{3}}\frac{1}{n}.
\label{NS}
\end{equation}
If the signals were monochromatic one would expect to resolve
roughly one source per frequency bin. Consequently, by equating
$N_s$ to 1 and assuming 1 year of observation time, we would get
an upper frequency cutoff equal to $f_o$. However, by accounting
in the data analysis for the modulation of the signal due to the
motion of LISA around the Sun, it is possible to resolve signals
of equal frequency but incoming from different sky positions.

To estimate this effect we represent the LISA response, to a
monochromatic gravitational wave signal, in the following form
\begin{equation}
h(t)=A_o\cos\Phi(t) \ ,
\end{equation}
where the amplitude $A_o$ is assumed to be constant, and the phase
$\Phi(t)$, which includes the Doppler modulation, is given by
\begin{equation}
\Phi(t) = \omega t + \frac{\omega R}{c}\cos(\lambda)\cos(\Omega t
- \beta - \phi_o)
        + \Phi_o.
\end{equation}
Here $\omega$ is the angular frequency of the wave, $\Omega =
2\pi/1$yr, ($\lambda, \beta$) are ecliptic coordinates of the
source, $R = 1$AU, $\phi_o$ is a known phase determining the
position of the detector in the orbit around the Sun and $\Phi_o$
is an unknown, constant phase of the signal. This is a reasonable
approximation since the time scale over which the detector's response
amplitude changes is significantly longer than that of the phase.

It is convenient to introduce the following linear parameterization of
the signal
\begin{equation}\label{LM}
h(t) = A_{o}\cos[\Phi_o + \omega t + A\cos(\Omega t) +
       B\sin(\Omega t)],
\end{equation}
where
\begin{eqnarray}
% \nonumber to remove numbering (before each equation)
  A &=& \frac{\omega R}{c}\cos\lambda\cos(\beta - \phi_o), \\
  B &=& \frac{\omega R}{c}\cos\lambda\sin(\beta - \phi_o).
\end{eqnarray}
In this parameterization the phase of the signal is a linear function
of the signal's unknown parameters ($\Phi_0$, $\omega$, A, B). The
reduced normalized correlation function of the signal is given by
\cite{JK00}
\begin{eqnarray}\label{CF}
\nonumber C(\Delta w,\Delta A, \Delta B) =
    &&(\int^1_0\cos[\Delta w x + \Delta A\cos(2\pi n x) + \Delta
    B\sin(2\pi n x)]\, dx)^2 + \\
    &&(\int^1_0\sin[\Delta w x + \Delta A\cos(2\pi n x) + \Delta
    B\sin(2\pi n x)]\, dx)^2 ,
\end{eqnarray}
where we have introduced a new dimensionless frequency parameter
$w = 1 {\rm yr} \ n \ \omega$, and the new dimensionless time
variable $x = t/({1 {\rm yr}} \ n)$.  The correlation function
above is derived under the assumption that, over the bandwidth of
the signal, the spectral density of the noise can be considered
constant. As a consequence of the linearity of the phase of the
signal in the parameters, the correlation function of two signals
depends only on the difference between the values of the
parameters and not on their absolute values. In the above
parameterization the parameter space can be visualized as a
truncated cone whose base and top are discs of radii
$\frac{\omega_u R }{c}$ and $\frac{\omega_l
  R}{c}$ respectively, with $\omega_u$ and $\omega_l$ being the
angular frequencies of the upper and the lower edge of the bandwidth
of the detector. For a given angular frequency $\omega$ the parameter
space is a disc of radius $\frac{\omega R }{c}$. Each point of the
disc corresponds to two positions in the sky which differ by the
sign in the ecliptic latitude angle $\lambda$.

In order to derive the background upper frequency cutoff we need to
estimate the number of signals resolvable in a given frequency bin.
This can be done by studying the correlation function between the
signals in the A-B plane, and choosing a particular value for it. In
this paper we have assumed two signals to be resolvable if their
normalized correlation is less than 1/2. This choice of course is not
optimal, and the final result of the upper frequency cutoff will
depend on it. However, larger values of the correlation function imply
a lower upper frequency cutoff, as it will be shown below.  Here we do
not identify the optimal value of the correlation function, which will
be estimated in a forthcoming paper via Monte Carlo simulations. The
reader should keep in mind this point when interpreting the
conclusions of this paper.

The correlation surface in the A-B plane can be approximated by an
ellipse whose area, $v_{AB}$, can be calculated using the covariance
matrix (which is the inverse of the Fisher information matrix). The
Fisher matrix, $\Gamma_{ij}$, for the signal (\ref{LM}) is given by
\begin{equation}\label{FM}
\Gamma_{ij} =
\left(%
\begin{array}{cccc}
  1                             & \frac{1}{2} & \frac{\sin(2n\pi)}{2n\pi}                               & \frac{1 - \cos(2n\pi)}{2n\pi}  \\
  \frac{1}{2}                   & \frac{1}{3} & -\frac{1 - \cos(2n\pi) - 2n\pi\sin(2n\pi)}{(2n\pi)^2}   & \frac{\sin(2n\pi) - 2n\pi\cos(2n\pi)}{(2n\pi)^2} \\
  \frac{\sin(2n\pi)}{2n\pi}     & -\frac{1 - \cos(2n\pi) - 2n\pi\sin(2n\pi)}{(2n\pi)^2}                 & \frac{4n\pi + \sin(4n\pi)}{8n\pi} & \frac{1 - \cos(4n\pi)}{8n\pi} \\
  \frac{1 - \cos(2n\pi)}{2n\pi} & \frac{\sin(2n\pi) - 2n\pi\cos(2n\pi)}{(2n\pi)^2}                      & \frac{1 - \cos(4n\pi)}{8n\pi} &  \frac{4n\pi - \sin(4n\pi)}{8n\pi} \\
\end{array}%
\right),
\end{equation}
where $i,j = (\Phi_o, w, A, B)$. Its expression has been derived
under the assumption that, over the bandwidth of the signal, the
spectral density of the noise can be considered constant. If we
also assume the integration time to be a integer multiple of 1
year, then the Fisher matrix becomes
\begin{equation}\label{FM1}
\Gamma_{ij} =
\left(%
\begin{array}{cccc}
  1   & 1/2 & 0   & 0 \\
  1/2 & 1/3 & 0   &-1/(2n\pi) \\
  0   & 0   & 1/2 & 0 \\
  0   & -1/(2n\pi) & 0 & 1/2 \\
\end{array}%
\right) \ .
\end{equation}
In the above expressions for $\Gamma_{ij}$ we have also normalized the
signal-to-noise ratio to 1.

Since the area of the correlation ellipse is given by
\begin{equation} v_{AB} = \frac{\pi\sqrt{det(C_2)}}{2} =
\frac{\pi^2n}{\sqrt{\pi^2n^2 - 6}},
\end{equation}
where $C_2$ is the 2 by 2 sub-matrix of the covariance matrix for the
parameters A and B, it follows that the number of resolved signals in
a frequency bin centered on the frequency $f$ is given by the ratio of
the area of the disc of radius $\frac{2\pi f R }{c}$, and the area of
the correlation ellipse $v_{AB}$. In mathematical terms we have:
\begin{equation}
N_r = N_o \left(\frac{f}{f_o}\right)^2,
\end{equation}
with
\begin{equation}
N_o = \frac{\pi(2\pi f_o R/c)^2}{v_{AB}}.
\end{equation}
If we equate the number of resolved signals, $N_r$, to the number $N_s$
of expected signals in one frequency bin (Eq. (\ref{NS})), we
get the following formula for the upper frequency cutoff, $f_r$, of
the background
\begin{equation}
f_r = \frac{f_o}{(N_o n)^{3/17}}. \label{FR}
\end{equation}
Equation (\ref{FR}) implies a frequency cutoff $f_r = 10^{-3.0} \ $Hz
when a year of observation time is assumed.

By using the expression of the number of signals per frequency bin
given in equation (\ref{DF}), we find that the number of detached
white dwarf binaries above the frequency $f_r$ is around $6.8 \
\times 10^4$. Although all these binaries can in principle be
resolved, not all of them can be detected as some will have an
amplitude smaller than the value of the detection threshold. To
calculate the number of detectable binaries we first need to
estimate the threshold corresponding to a satisfactory confidence
of detection.  We assume that we shall process the data by matched
filtering. For the linear phase model of the signal given above we
can adopt methods developed in \cite{JKS98}. First we need to
calculate the number of cells, $N_c$, in the parameter space.
These are the number of realizations of the optimal filter whose
correlation is smaller than a pre-chosen value (1/2 in our case).
This number is given by the ratio between the volume of the
characteristic correlation hypersurface and the volume of the
parameter space over which the search is performed. In our case
the volume of the parameter space is given by
\begin{equation}
V = n \ {\rm yr} \ \frac{\pi}{3}\left(\frac{1 \mbox{AU}}{c}\right)^2 \
(\omega_u^3 - \omega_l^3) \ ,
\end{equation}
while the volume $v$ of the correlation ellipsoid can be obtained by
using the covariance matrix:
\begin{equation}\label{vol}
    v = \frac{\pi\sqrt{2 \ det(C_3)}}{3} \ .
\end{equation}
Here $C_3$ is the 3 by 3 sub-matrix of the covariance matrix for the
parameters $w, A$, and $B$. In reference \cite{JKS98}, Eq.(77), it was
shown that for Gaussian noise, the false alarm probability can be
written as follows
\begin{equation}
P_{\rm F} = 1 - [1 - e^{-{\cal F}_0}]^{N_c} \ ,
\label{son}
\end{equation}
where ${\cal F}_0$ is the threshold on the following optimal
statistics function \cite{JKS98}
\begin{equation}
{\cal F} = \frac{1}{n \ {\rm yr} \ S_c} \left|\int_0^{n {\rm yr}}
\ y(t) \ e^{\ i \ [ A\cos(\Omega t) + B\sin(\Omega t) + \omega t
]} \ dt\right|^2 . \label{gr2}
\end{equation}
In Eq. (\ref{gr2}) $S_c$ is the value of the two-sided power
spectral density of the noise estimated in the center of the
frequency bandwidth of the signal, and $y (t)$ is the LISA data
stream.

Equation (\ref{son}) implies that, with a false alarm probability of
$1$ percent, the corresponding threshold signal-to-noise ratio is
equal to $7.6$.  The corresponding number of resolvable white dwarf
binaries with signal-to-noise above this threshold is equal to $3365$,
under the assumption of sources uniformly distributed in the galactic
disk. As a comparison, the number of resolvable binaries calculated
without including the effects of the phase modulation of the signals
is equal to 919, a factor of $3.7$ smaller than what we estimate.

This threshold is of course rather conservative, since it reflects
the assumption that we are searching for one or more signals in
the data, and we should therefore regard it as an upper-bound. In
reality these signals are present in the LISA data and, in order
to calculate the lower bound for the threshold, we can assume each
cell to contain a signal. Consequently, the false-alarm
probability becomes
\begin{equation}
P_{\rm F} = e^{-{\cal F}_0} \ ,
\label{son1}
\end{equation}
since now $N_c = 1$ in equation (\ref{son}). The corresponding
threshold signal-to-noise ratio goes down to 2.7, again for a
false alarm probability of $1$ percent. This implies that $47440$
white-dwarf binaries of signal-to-noise ratio larger than this
threshold will be resolvable. If we neglect the effects of the
phase modulation of the signals, the number of resolvable binaries
is equal to $13178$ instead.

As a demonstration of how well the parameters of the signals can
be estimated by using the maximum likelihood method, we have
performed a numerical simulation of our technique. We first
estimate the parameters of the strongest signal, we remove it from
the data, and then iterate these two steps until no signal crosses
our predefined threshold.  The numerical procedure for locating
the maximum consists of two steps. The first is a coarse search of
the maximum of the likelihood function on a predefined grid in the
parameter space.  This is then followed by a refinement around the
region of the parameter space where the maximum identified by the
coarse search is located.  This second stage of the maximization
is performed by using a numerical implementation of the
Nelder-Mead algorithm, where the starting point of the
maximization is determined by the values of the parameters
identified by the coarse search.

A graphical representation of the effectiveness of our method is
presented in Figures 1 and 2, which correspond to two different
combinations of signals in the LISA data. Both figures show two
sinusoidal signals of identical frequencies, and incoming from
different two directions in the sky. The correlation between the
two signals was chosen to be approximately equal to 1/2. In figure
1 we input signals with signal-to-noise ratios equal to 7 and 20,
while in Figure 2 the signal-to-noise ratios have been increased
to 20 and 60 respectively.  The common frequency of the signals
($f=1.5 \ \times 10^{-3}$ Hz), and their incoming directions in
the sky are unchanged in both figures.

\begin{figure}
  % Requires \usepackage{graphicx}
  \includegraphics[width=13cm]{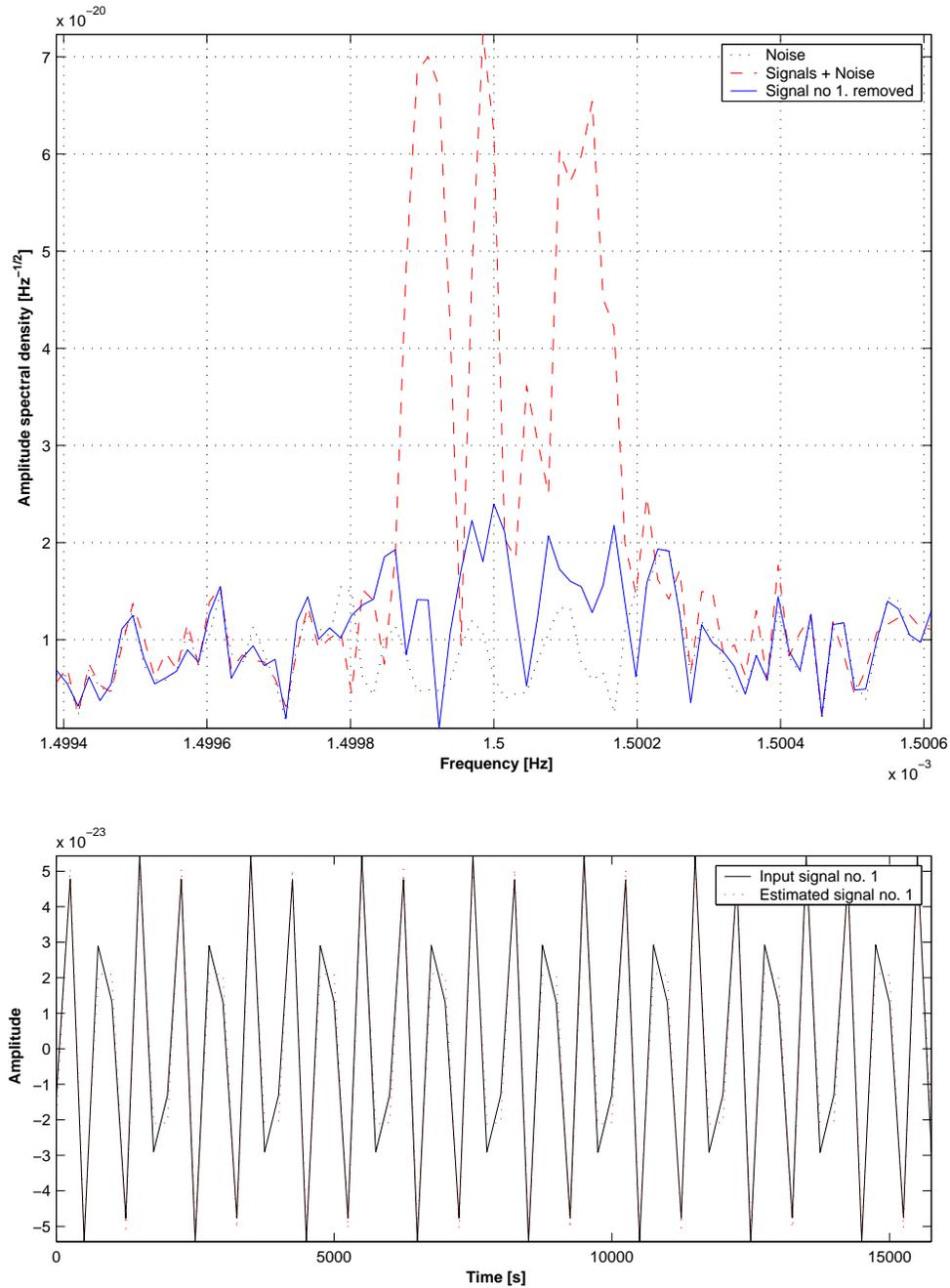}\\
  \caption{Estimation and removal of one of two signals from the LISA data stream
    using a two-step maximum likelihood estimation. The two signals
    have signal-to-noise ratios equal to 7 and 20 respectively, their
    frequencies are equal, but they originate from different points in
    the sky. The upper part of the figure shows the power spectra of
    the signals and the noise (dashed line), of the data after the
    stronger signal is removed (solid line), and of the noise only (dotted line).
    The lower part instead compares
    the shape of the signal injected (solid line) into the LISA data against that
    of the estimated one (dotted line).}
\label{sim_20_7a}
\end{figure}

\begin{figure}
  %Requires \usepackage{graphicx}
  \includegraphics[width=13cm]{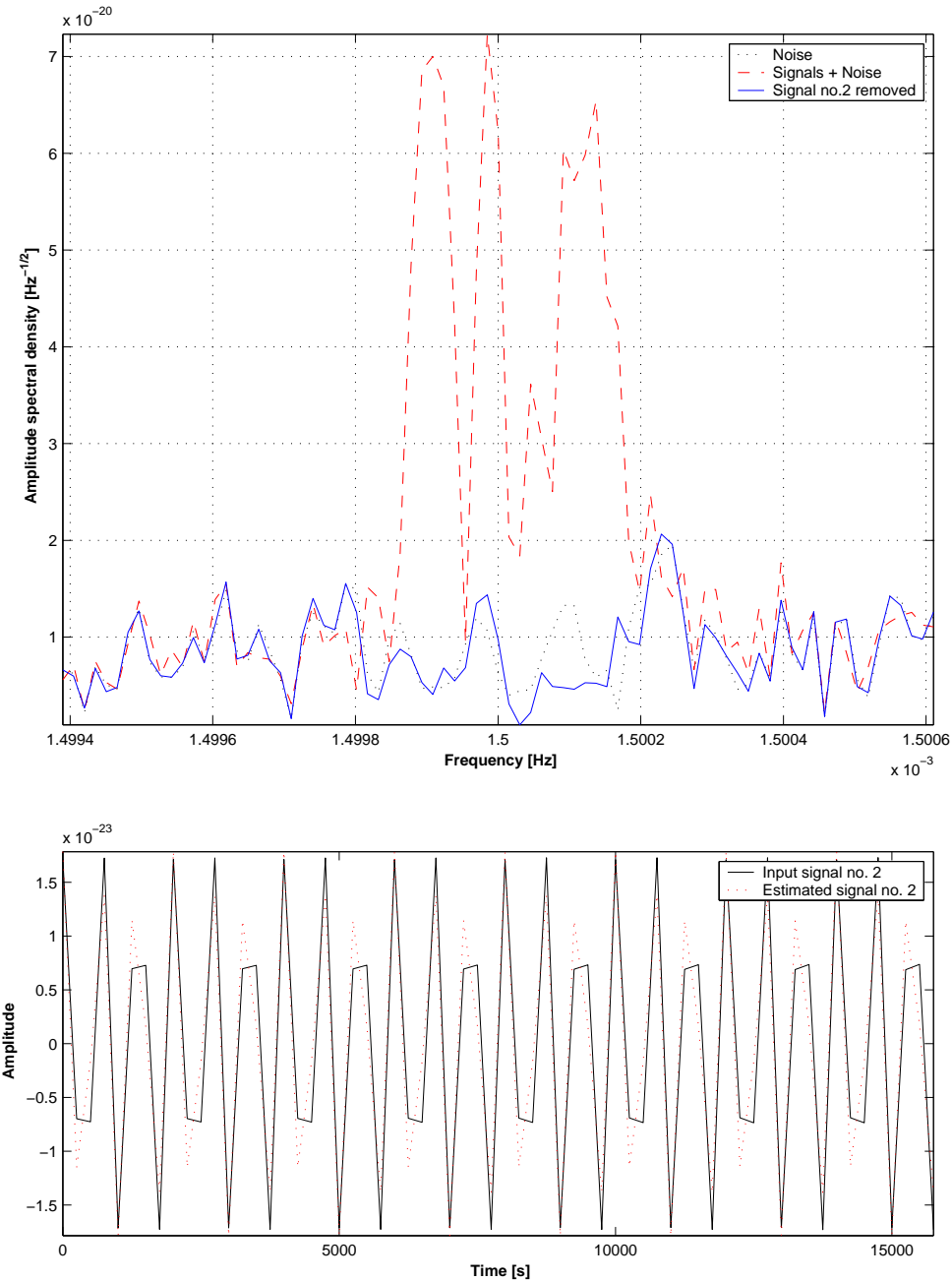}\\
  \caption{The signals are as in Figure 1a, but now the spectra shown
    include one in which the signals are both
    removed (solid line).}
\label{sim_20_7b}
\end{figure}

\begin{figure}
  % Requires \usepackage{graphicx}
  \includegraphics[width=13cm]{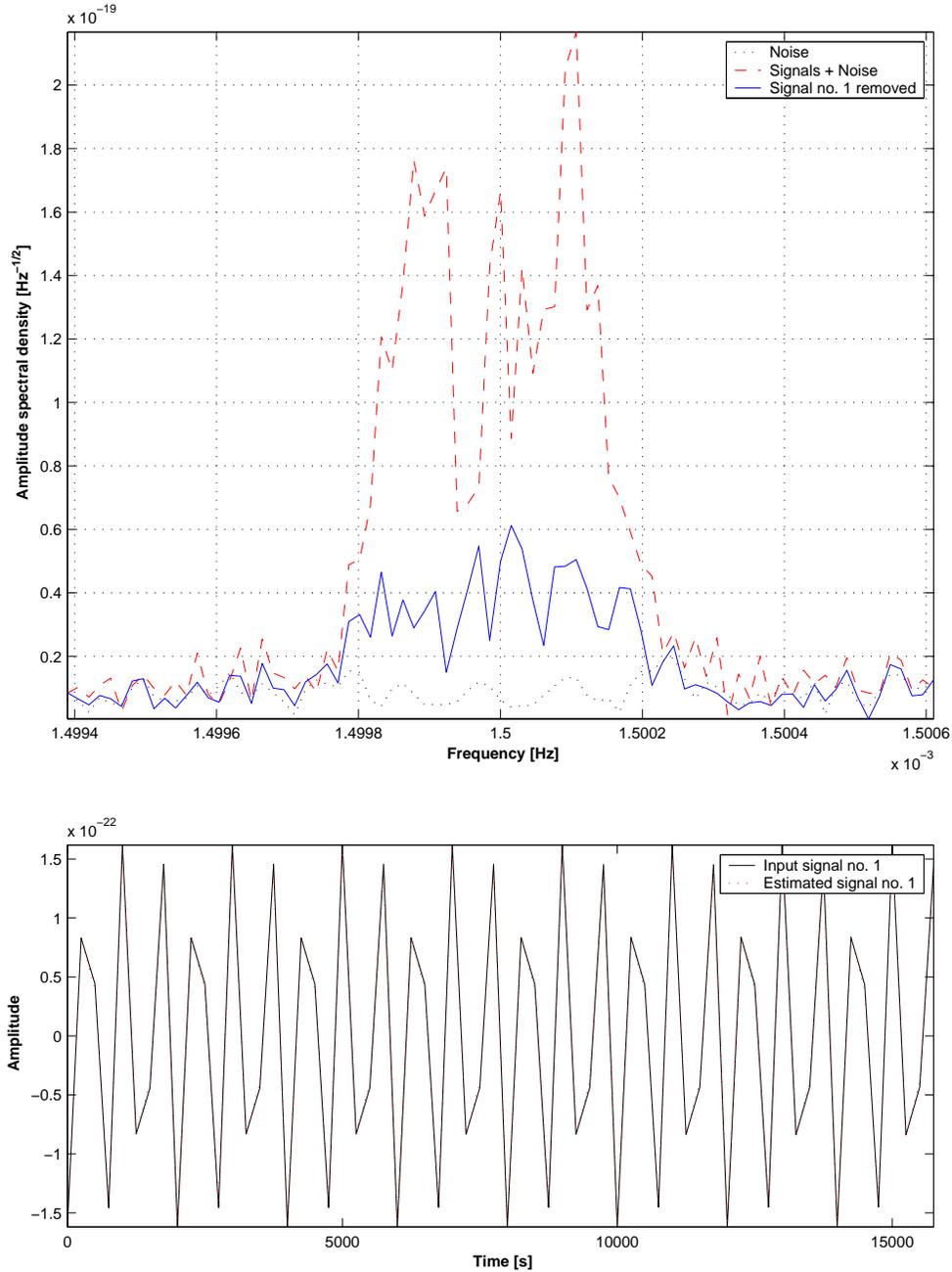}\\
  \caption{As Figure 1a, but now the signal-to-noise ratios have been
    increased to 20 and 60 respectively}
\label{sim_60_20a}
\end{figure}

\begin{figure}
  % Requires \usepackage{graphicx}
  \includegraphics[width=13cm]{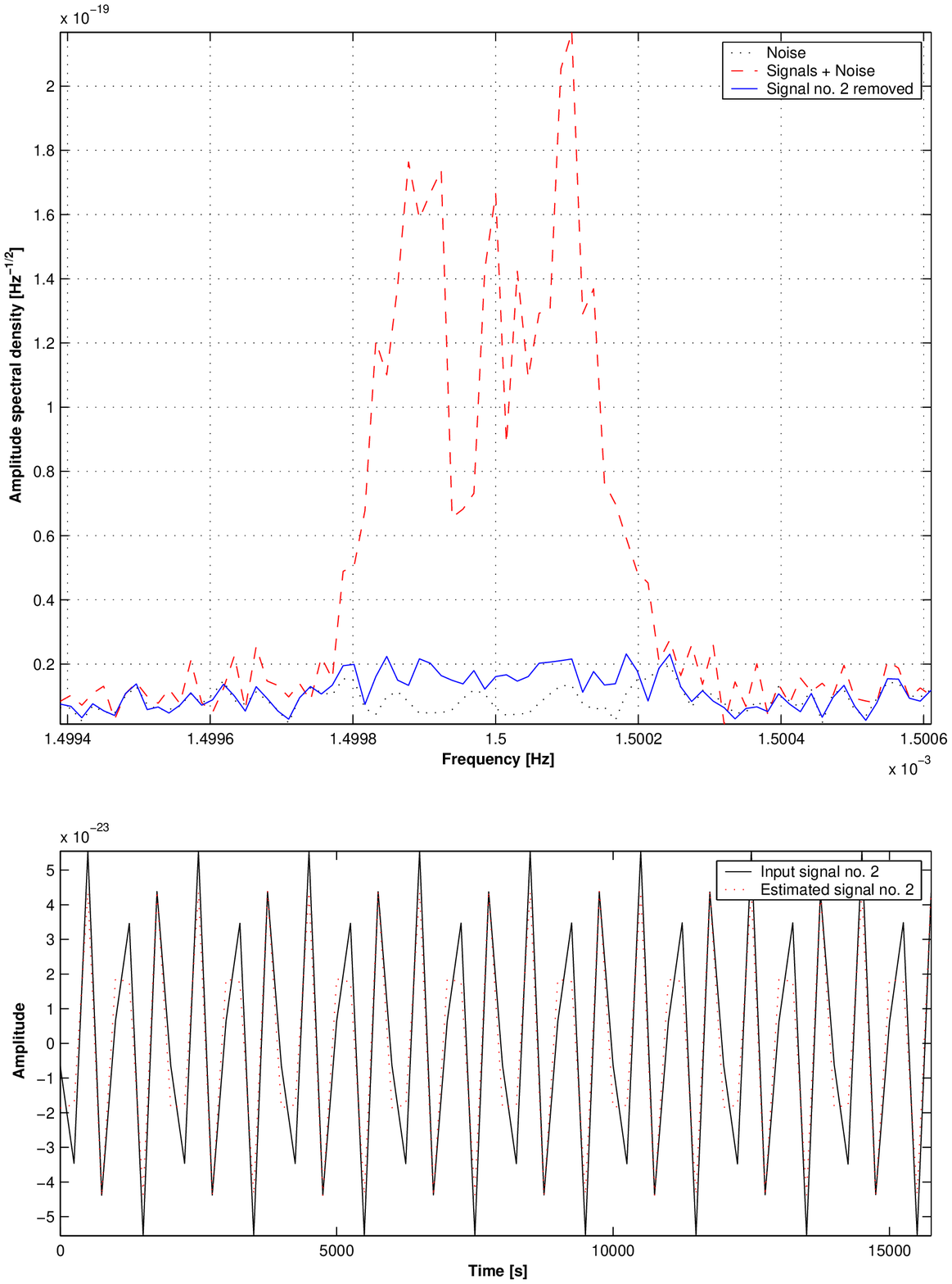}\\
  \caption{The same as Figures 1b, but the signals have signal-to-noise
  ratios equal to 20 and 60}\label{sim_60_20b}
\end{figure}

A visual inspection of the plots shows that the estimation of the
parameters and the resolution of the signals (with the use of the
maximum likelihood method) can be expected to be satisfactory for
removing them from the data. This will make possible the search of
other signals potentially present in the LISA data.

\section*{Acknowledgments}

One of us (AK) acknowledges support from the National Research
Council under the Resident Research Associateship program at the
Jet Propulsion Laboratory, California Institute of Technology, and
from the Polish State Committee for Scientific Research (KBN)
through Grant No. 2P03B 094 17. The research was performed at the
Jet Propulsion Laboratory, California Institute of Technology,
under contract with the National Aeronautics and Space
Administration.

\end{document}